\documentclass[a4paper,fleqn,usenatbib]{mnras}
\usepackage{newtxtext,newtxmath}
\usepackage[T1]{fontenc}
\usepackage{ae,aecompl}
\usepackage{multirow}
\usepackage{graphicx}    
\usepackage{amsmath}    
\usepackage{ulem}
\usepackage{hyperref}
\usepackage[export]{adjustbox}
\usepackage{pgffor}

\def\araa{ARA\&A}%% Annual Review of Astron and Astrophys
\def\apj{ApJ}%% Astrophysical Journal
\def\apjl{ApJ}%% Astrophysical Journal, Letters
\def\apjs{ApJS}%% Astrophysical Journal, Supplement
\def\ao{Appl.~Opt.}%% Applied Optics
\def\aap{A\&A}%% Astronomy and Astrophysics
\def\jcomputphys{J. Comput. Phys} %% Journal of Computational Physics
\def\mnras{MNRAS}%% Monthly Notices of the RAS
\def\philtransrsoca{Phil. Trans. R. Soc. A} 
\def\solphys{Sol.~Phys.}%% Solar Physics
\hypersetup{draft}

\newcommand{\ttres}{60}
\newcommand{\tcuatro}{80}

\newcommand{\boltzmann}{k_\mathrm{B}}
\newcommand{\protonmass}{m_\mathrm{p}}

\title[]{Thermal conduction effects on formation of chromospheric solar tadpole-like jets}

\author[Navarro et al]{
Anamar\'ia Navarro,$^1$ \thanks{E-mail:ana.navarro1@correo.uis.edu.co}
F. D. Lora-Clavijo,$^1$  
K. Murawski$^2$  and 
Stefaan Poedts$^3$ \\
$^{1}$Grupo de Investigaci\'on en Relatividad y Gravitaci\'on, Escuela de f\'isica, 
Universidad Industrial de Santander,\\ A.A. 678, Bucaramanga, Colombia \\
$^2$Institute of Physics, University of Maria Curie-Sk\l{}odowska, Pl. M. Curie-Skłodowskiej 5, 20-031 Lublin, Poland \\
$^3$Centre for mathematical Plasma Astrophysics, Dept. of Mathematics, KU Leuven, Celestijnenlaan 200B, 3001 Leuven, Belgium}

\begin{document}

\maketitle

\begin{abstract}
We measure the effects of non-isotropic thermal conduction on  generation of solar chromospheric jets through numerical simulations carried out with the use of one fluid MHD code MAGNUS. Following the work of  \cite{pseudoshoks_nature_2018}, we consider the atmospheric state with a realistic temperature model and generate the ejection of plasma through a gas pressure driver operating in the top chromosphere. We consider the magnetic field mimicking a flux tube and perform parametric studies by varying the magnetic field strength and the amplitude of the  driver. We find that in the case of thermal conduction the triggered jets exhibit a considerably larger energy and mass fluxes and their shapes are more collimated and penetrate more the solar corona than for the ideal MHD equations. Low magnetic fields allow these jets to be more energetic, and larger magnetic fields decrease the enhancement of mass and energy due to the inclusion of the thermal conductivity.
\end{abstract}


% &&&&&&&&&&&&&&&&&&&&&&&&&&&&&&&&&&&&&&&&&&&&&&&&
\section{Introduction}
% &&&&&&&&&&&&&&&&&&&&&&&&&&&&&&&&&&&&&&&&&&&&&&&&

Revealing the heating mechanism of the solar corona is one of the most important research inquiries in solar physics. This mechanism should compensate the thermal conduction, radiation and solar wind losses estimated by \cite{Withbroe_Noyes_1977}. It is believed that the origin of high temperatures in the corona is of magnetic nature and that the energy source lies in the solar surface plasma motions \citep{Arregui2015}. Several mechanisms have been considered to contribute to the heating like the dissipation of magnetic energy by magnetic reconnection \citep{1986ApJ...311.1001V}, current cascades \citep{1963IAUS...16...11P}, viscous turbulence \citep{1986ApJ...311.1001V}, magnetic braiding \citep{2004ApJ...617L..85P}, the dissipation of wave energy \citep{1947MNRAS.107..211A, 1978ApJ...226..650I, 1983A&A...117..220H, 1991assm.conf..137G, 1997A&A...324...11H}, and the mass flow cycle between the chromosphere and corona \citep{2012ApJ...749...60M}. 

Recently, \cite{pseudoshoks_nature_2018} reported a new mechanism, a jet with rarefied plasma, which was observed using the Interface Region Imaging Spectrograph on 8 October 2014 \citep{2014SoPh..289.2733D}. Numerical simulations trying to reproduce this phenomenon showed that it could carry an energy of $\sim 10^3$ W m$^{-2}$, which is in agreement with the values estimated by \cite{Withbroe_Noyes_1977}. Due to the high energy transported by it, this interesting new phenomenon could make an important contribution to the coronal heating problem. These 2D simulations were performed using the JOANNA code \citep{2018MNRAS.481..262W}, solving two fluid equations (for ions + electrons and neutrals) under ideal conditions, that is, without considering non-adiabatic terms. However, in these specific conditions the thermal conduction should be considered because of the steep gradient temperatures in the transition region, and due to the shape of the magnetic field that conducts very efficiently the heat flux along its field lines \citep{Spitzer1956}.   Additionally, it is necessary to determine the influence of the magnetic field strength in the phenomena involved in the ejections of plasma, since some observations suggests that the magnetic fields in coronal loops seem to be much larger than traditionally considered \citep{Kuridze_2019}.

In this paper we simulate the ejection of a tadpole jet generated in the chromosphere and study the influence of the thermal conduction and the magnetic field amplitude on its dynamic. The numerical simulations are carried out with the MAGNUS code \citep{magnus} and the model of the gas pressure driver is taken from the work of \cite{pseudoshoks_nature_2018}. The magnetic field modelled corresponds to a flux tube and the temperature  is taken from the semi-empiric profile of \cite{2008AvrettLoeser}. We perform parametric studies by varying the amplitude of the driver, and the magnetic field strength of the flux tube and repeat the set of simulations under non-adiabatic conditions.  This paper is organized as follows. In Section~\ref{sec_numerical_model}, there is a description of the numerical model, the MHD equations with heat conduction, the thermal conduction model, the initial equilibrium state and the boundary conditions. In Section \ref{sec_results}, we present the comparison of the mass density for a simulation with thermal conduction and another one without, the vertical velocity, the gas pressure, the Mach number and the heat loss/gain rate for a given time. We present the comparison between the energy and mass flux carried by the jet for all the simulations and also the horizontally averaged heat loss/gain rate at a given time for different set of parameters. Finally, in Section~\ref{sec_conclusions} we draw our conclusions.

% &&&&&&&&&&&&&&&&&&&&&&&&&&&&&&&&&&&&&&&&&&&&&&&&
\section{Numerical model}\label{sec_numerical_model} 
% &&&&&&&&&&&&&&&&&&&&&&&&&&&&&&&&&&&&&&&&&&&&&&&

We model dynamics of the plasma of the solar atmosphere by MHD equations, written in a conservative form as 
\begin{eqnarray}
\partial_t \vec{U} + \partial_i \vec{F^i} = \vec{S} \, . \label{MHD_eq_ConservativeForm}
\end{eqnarray}
Here $\vec{U}$ is a state vector, $\vec{F^i}$ is the flux along $i-$axis and $\vec{S}$ is a source vector term, all are given as
\begin{eqnarray}
& & \vec{U} = \left[ \begin{array}{c}
\varrho \\
\varrho \vec{v} \\
 E \\
\vec{B} 
\end{array} 
  \right] \, ,  \hspace{3mm}  
\vec{F^i} = \left[ \begin{array}{c} \varrho v^i \\  
 \varrho v^i v_j - \frac{B^i B_j}{\mu_\mathrm{0}} + p_\mathrm{T}\delta^i_j \\
( E + p_\mathrm{T})v^i - \frac{B^i(\vec{B}\cdot \vec{v})}{\mu_\mathrm{0}} \\
v^i B_k - v_k B^i   \end{array}  \right] \, , \qquad  
\end{eqnarray} \begin{eqnarray}
 \vec{S} = \left[ \begin{array}{c}
0 \\
- \varrho \vec{g} \\ 
\varrho \vec{v}\cdot \vec{g} - \nabla \cdot \vec{q}   \\
0
 \end{array} \right] \, ,
\end{eqnarray}
where $\varrho$ denotes mass density, $\vec{v}$ the velocity, $\vec{B}$ the magnetic field, $p$ the gas pressure, $\mu_\mathrm{0}$ the magnetic permeability of free space,   $p_\mathrm{T} = p + B^2/2\mu_\mathrm{0}$ the total gas pressure,  $\vec{q}$ the heat flux, $\vec{g} = [0,0,-g]$ the gravitational acceleration with its magnitude of $274\;$m s$^{-2}$, and the total energy density  
\begin{eqnarray}
E = \frac{\varrho v^2}{2} + \frac{p}{\Gamma-1} + \frac{B^2}{2 \mu_0} \, . 
\end{eqnarray}
 The fluid obeys the ideal gas law with adiabatic index $\Gamma = 5/3$.

The non-isotropic thermal conduction is operated along the magnetic field lines according to the classical model for magnetized plasma \citep{Spitzer2006}, leading to
\begin{eqnarray}
\vec{q} = - \frac{ \kappa T^{5/2} ( \vec{B} \cdot \nabla T ) \vec{B} }{B^2} \, , 
\end{eqnarray}
where $\nabla T$ is the temperature gradient and $\kappa$ is the thermal conductivity coefficient $\kappa = 10^{-11}$ W m$^{-1}\;$K$^{-7/2}$.

\begin{figure}
\centering
\includegraphics[width = 0.35\textwidth]{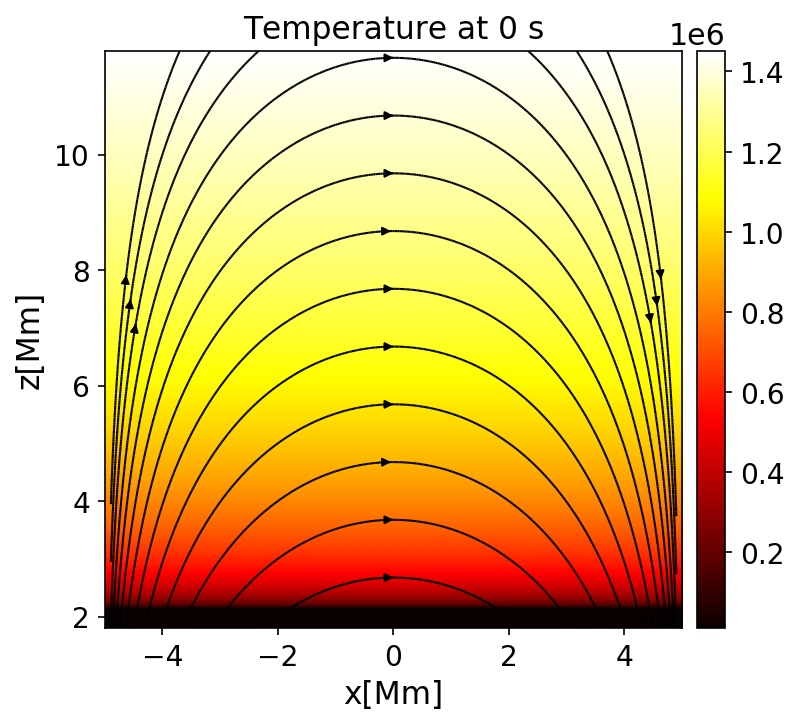} 
\caption{Colour map of the temperature profile and the configuration of the initial magnetic field lines. \label{initial_profiles1}}
\end{figure}

% &&&&&&&&&&&&&&&&&&&&&&&&&&&&&&&&&&&&&&&&&&&&&&&&
\subsection{Initial data} \label{subsec_initial_data} 
% &&&&&&&&&&&&&&&&&&&&&&&&&&&&&&&&&&&&&&&&&&&&&&&
Following the work of \cite{pseudoshoks_nature_2018}, the hydrostatic state is obtained from the semi-empirical model for the temperature developed by \cite{2008AvrettLoeser}. Accordingly, from the hydrostatic equation we have
\begin{eqnarray}
\varrho_\mathrm{h} g + \frac{\partial p_\mathrm{h}}{\partial z} = 0 \; , \label{eq_equilibrium}
\end{eqnarray}
where the subscript $h$ denotes the hydrostatic value. Then,  the gas pressure can be specified from the ideal gas law,
\begin{eqnarray}
p_\mathrm{h} =  \frac{\boltzmann}{\protonmass \bar{m} } \varrho_\mathrm{h} T_\mathrm{h} \, . \label{eq_EOS}
\end{eqnarray}
Here $\protonmass$ corresponds to the proton mass, $\bar{m}$ is the mean mass, set to 1.24, and $\boltzmann$ denotes the Boltzmann's constant. In this way, the hydrostatic mass density and gas pressure are derived once  the temperature profile $T(z)$ is assumed, 
\begin{eqnarray}
& & p_\mathrm{h}(z) =  p_\mathrm{0} \exp{\left( -\frac{\protonmass \bar{m} g}{\boltzmann} \int_{z_\mathrm{0}}^z \frac{\mathrm{d}\tilde{z}}{T(\tilde{z})}  \right)} \, , \qquad  \\
& & \varrho_\mathrm{h}(z) = \frac{\protonmass \bar{m} }{\boltzmann} \frac{p_\mathrm{h}(z)}{T_\mathrm{h}(z)}  \, ,
\end{eqnarray}
where $z_\mathrm{0} = 10\;$Mm is the reference height.

The magnetic field, that corresponds to a flux tube can be found by a standard potential field solution of the static 2D MHD equations and it is given as \citep{Priest1982}  
\begin{eqnarray}
& & B_x = B_0 \cos\left( \frac{\pi x}{10}\right) \exp\left(- \frac{\pi z}{10}\right) \, ,\\
& & B_z = -B_0 \sin\left(\frac{\pi x}{10}\right)\exp\left(-\frac{\pi z}{10}\right) \, , 
\end{eqnarray}
where $B_0$ is the photospheric field magnitude which is varied in the simulation. Figure \ref{initial_profiles1} displays the initial configuration of the magnetic field lines and the plasma temperature. %The magnetic field modelled represents a 2D magnetic coronal arcade, which potential field models a system of neighbouring loops.
The temperature is $1 \times 10^4\;$K in the top of the chromosphere $(z=2.0 $ Mm) and in the transition region at $z = 2.1$ Mm grows abruptly, and smoothly increases to
$1.4 \times 10^6\;$K at $z = 12.0\;$Mm.

\begin{figure*}
\centering

\includegraphics[height = 0.18\textheight]{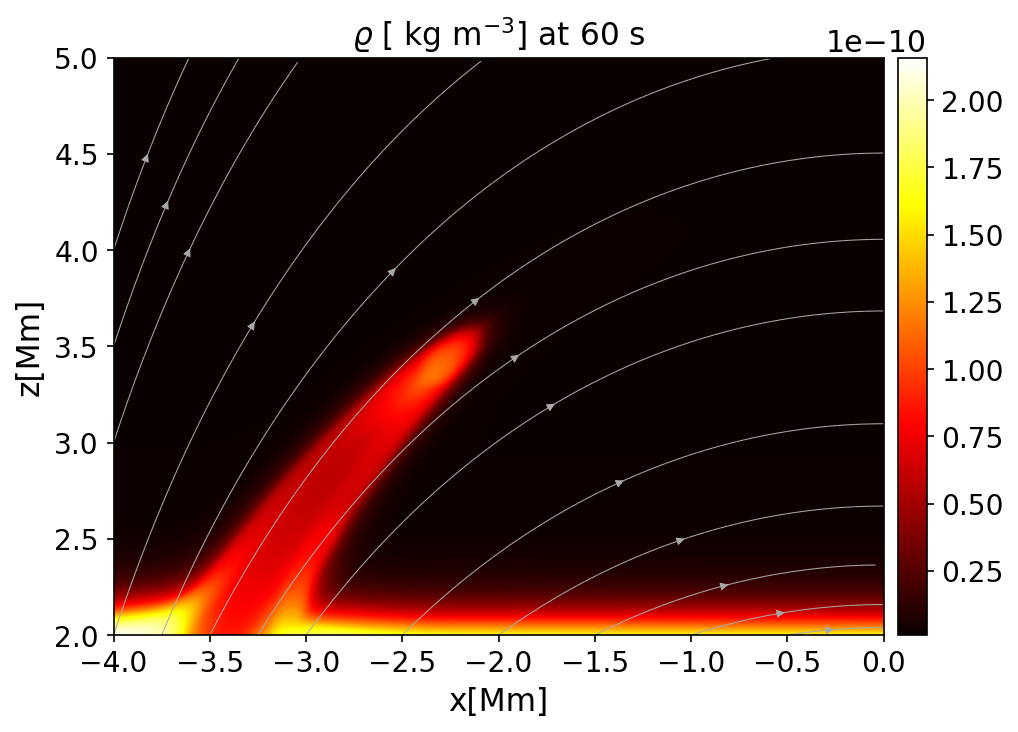}
\includegraphics[height = 0.18\textheight]{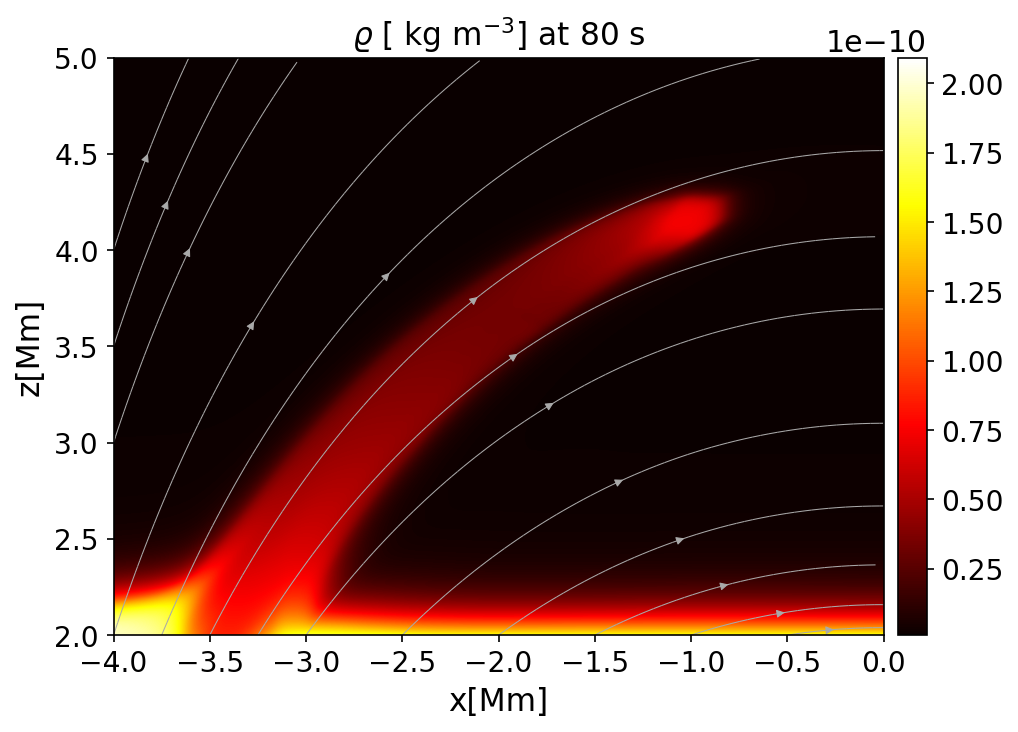}
\\
\includegraphics[height = 0.18\textheight]{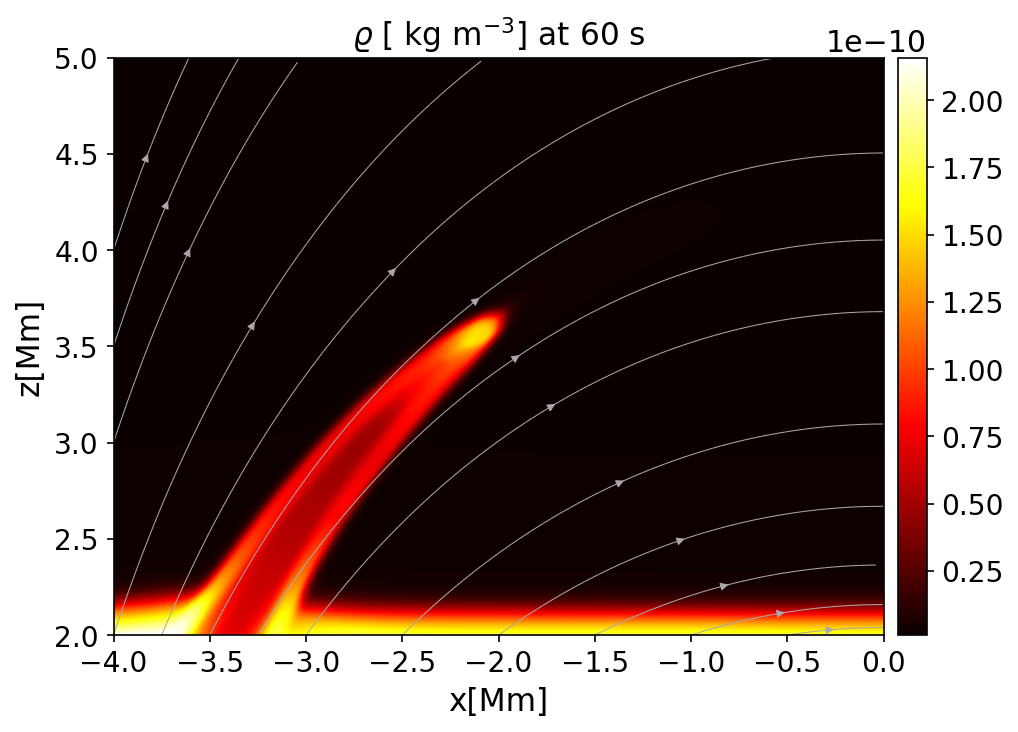}
\includegraphics[height = 0.18\textheight]{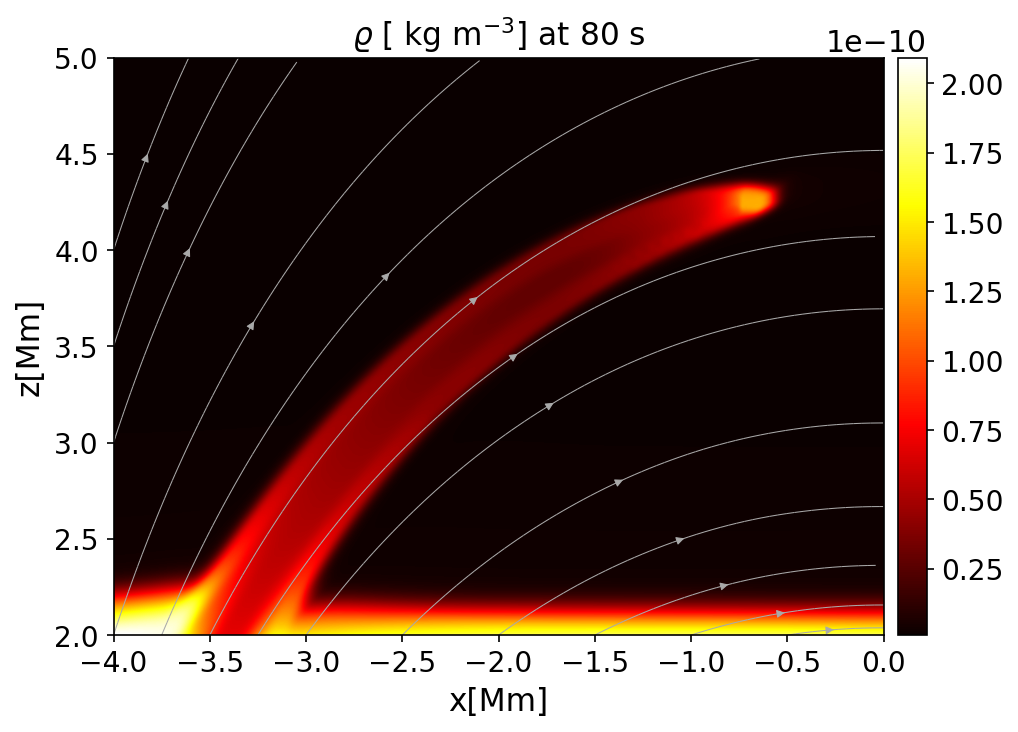}

\caption{Colour maps of mass density and magnetic field lines at $t = 60\;$s (left panels) s and $t = \tcuatro\;$s (right panels) for $A_p = $ 40, $B_0 =  40\;$G, adiabatic (top panels) and thermal conduction (bottom panels) cases. \label{fig_rho_linesB} }
\end{figure*}

\begin{figure*}
\centering
\includegraphics[height = 0.16\textheight] {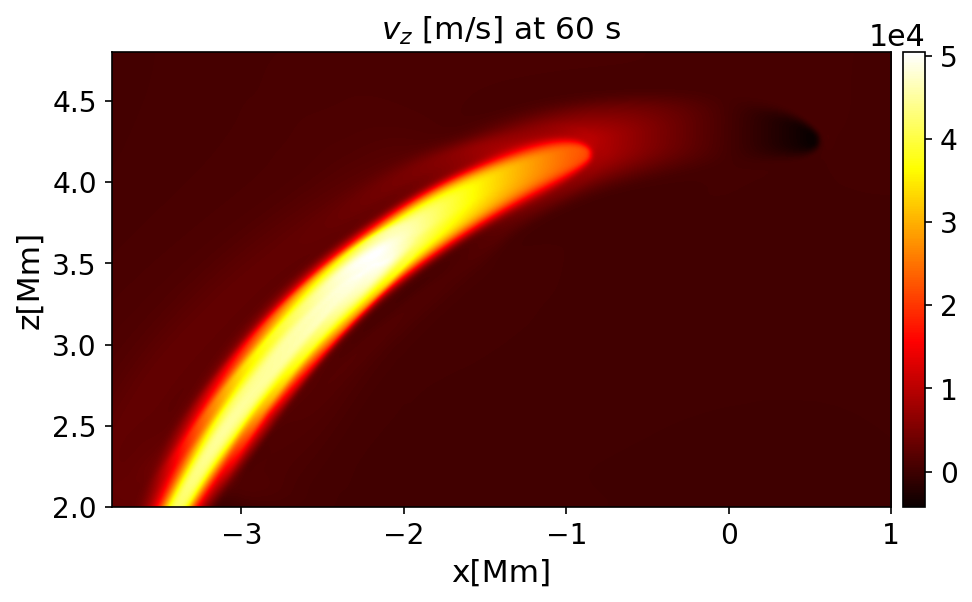}
\includegraphics[height = 0.16\textheight]{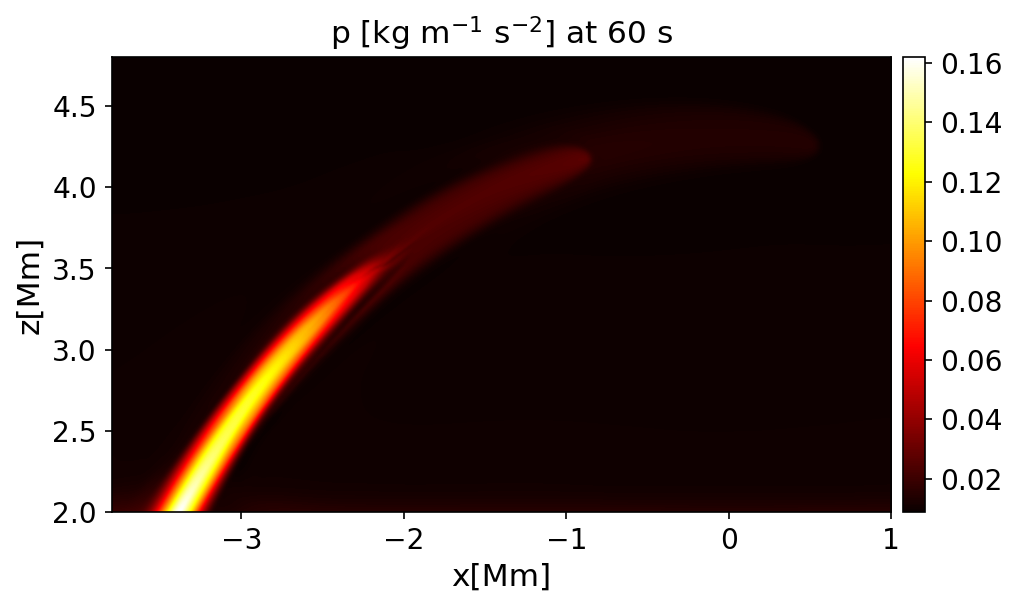}
\includegraphics[height = 0.16\textheight]{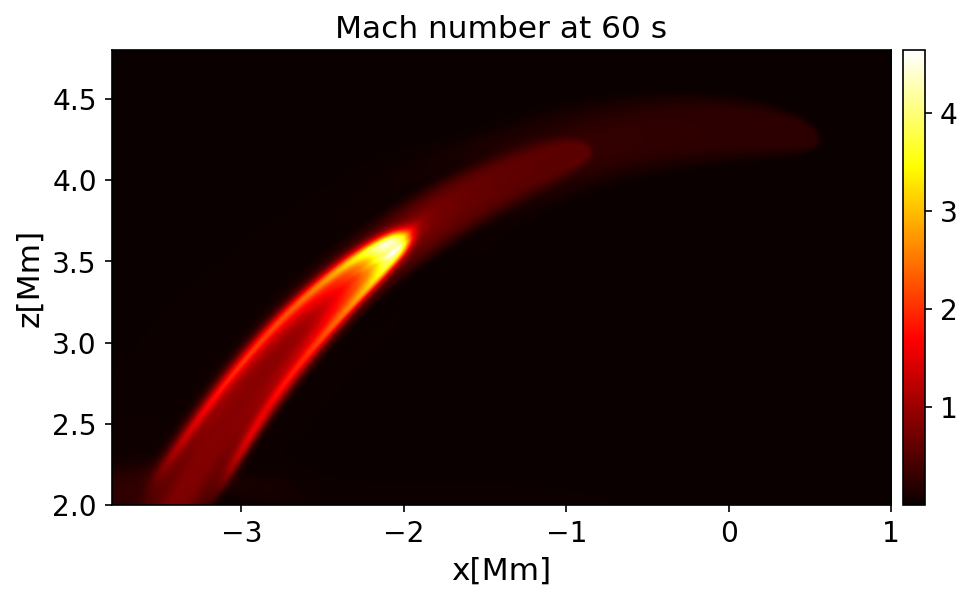}
\includegraphics[height = 0.16\textheight]{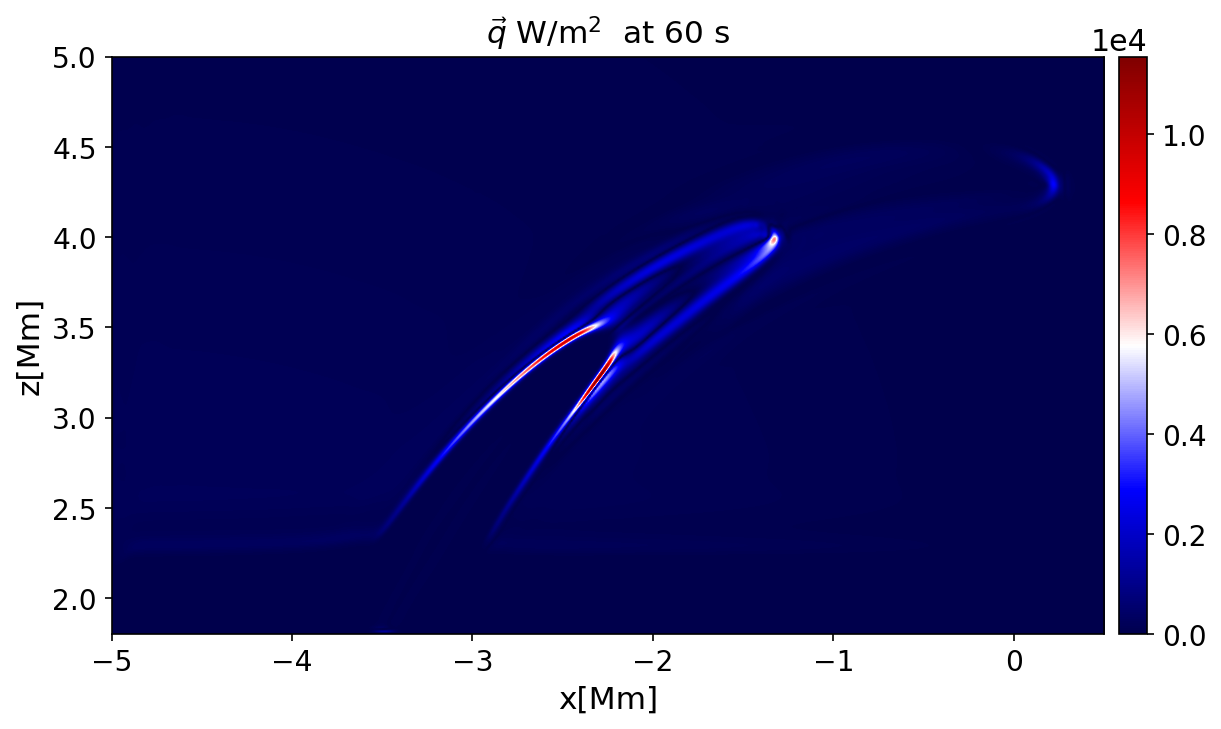}
\caption{Colour maps of the vertical velocity $v_z$ (left-top), gas pressure (right-top), Mach number (left-bottom) and the magnitude of the thermal conduction flux (right-bottom) at t = $60\;$s, for the simulation with thermal conduction, using $A_p = $ 40 and $B_0 = 40\;$G. \label{fig_2} }
\end{figure*}

\subsection{Simulation box and boundary conditions}\label{subsec_simulationbox} 

The equations are solved with the MAGNUS code \citep{magnus} in a simulation box of uniform grid with a 5 km resolution in both $x-$ and $z-$directions which covers the region of $[-5.0, 5.0]\times[1.8, 11.8]\;$Mm$^2$. The Courant-Friedrichs-Lewy (CFL) number is set to 0.1, a small number in order to obtain a small time step to solve the elliptic term introduced by the thermal conduction, and an adaptive time step. We use the HLLE Riemann solver \citep{Harten_etal_1983}, the van Leer slope limiter \citep{Vleer1977}, and a third order total variation diminishing Runge-Kutta \citep{Abramowitz1972}. The Flux Constrained Transport method \citep{CT_evans, CT_Balsara} is adapted to avoid the growth in time of $\nabla\cdot \vec{B}$.

The implemented boundary conditions are: outflow at the lateral sides and fixed to the initial plasma quantities at the bottom and topsides. By this, we mean that at the bottom and topsides we reset the values of the conservative variables to the initial state. Additionally,  a driver in the gas pressure is used at the bottom boundary and it is given by
\begin{eqnarray}
 p(x,t, z=z_0) = p_h(z=z_0) \left[ 1 + A_p \exp\left( -\frac{(x-x_0)^2}{w^2} \right)f(t)  \right]  \, , \\
 f(t) = \left\lbrace  \begin{array}{cl}
1 - \exp\left(-\frac{t}{\tau}\right) \, , & \text{for} \quad  t \leq \tau_\mathrm{max} \, ,  \\
\exp(-(t-\tau_\mathrm{max})/\tau) \, , & \text{for} \quad t > \tau_\mathrm{max} \, , 
\end{array} \right.
\end{eqnarray}
where $x_0 =  -3.4\;$Mm, $w = 0.1\;$Mm and $z_0 = 1.8\;$Mm.  Temporal profile f(t) is written with the intent to mimic a typical time profile of X-ray emission of solar flares \citep{2015ApJ...813...70F}. Here $\tau = 50\;$s and $\tau_\mathrm{max} = 30\;$s.

% &&&&&&&&&&&&&&&&&&&&&&&&&&&&&&&&&&&&&&&&&&&&&&&&&&&
\section{Numerical Results}\label{sec_results}
% &&&&&&&&&&&&&&&&&&&&&&&&&&&&&&&&&&&&&&&&&&&&&&&&&&&

In order to determine the effects of the thermal conduction on the jet generation and evolution, we ran two identical simulations that only differ by the dissipative term ($\nabla\cdot \vec{q}$) in the energy equation. Figure \ref{fig_rho_linesB} displays the density as a colour map and the magnetic field lines at $t = \ttres\;$s (left panels) and $t = \tcuatro\;$s (right panels). The top row corresponds to the adiabatic simulations while the bottom panels to  thermal dissipation switched on. We found that the simulations with thermal conduction narrows the ejected plasma and accelerates it more, going further into the corona than in the adiabatic case. Since the same colour scale is used to compare the results, it can be seen that in the adiabatic case the mass transported by the jet is more dissipated and with lower intensity. It is worth mentioning that the energy input through the boundaries is the same for each pair of simulations with and without thermal conduction, this is guaranteed by the boundary conditions applied.

In Figure \ref{fig_2} are plotted at $t=60\; $s   the vertical velocity, the gas pressure, the Mach number and the heat loss rate, respectively from top-left to bottom-right panels. They all correspond to the simulation with thermal conductivity, $A_p = 40$ and $B_0=40\;$G. In Figure \ref{fig_2}  the jet reached a height of $z = 3.6\;$Mm, in this region we detect a contact wave corresponding to the apex of the jet travelling with supersonic speed comparing  to the Mach number, since the velocity and gas pressure are not discontinuous, only the mass density. A supersonic slow shock can be seen at $x=$0.5 Mm, $z=4\;$Mm, where the plots illustrate a discontinuity in the velocity, gas pressure and mass density. The magnitude of the thermal conduction flux $|\vec{q}|$ is of the order of $1\times 10^4$ Wm$^{-2}$, which is in agreement with the losses for thermal conduction estimated by \cite{Withbroe_Noyes_1977} for active regions.  

\begin{figure}
\centering
\includegraphics[width = 0.32\textwidth]{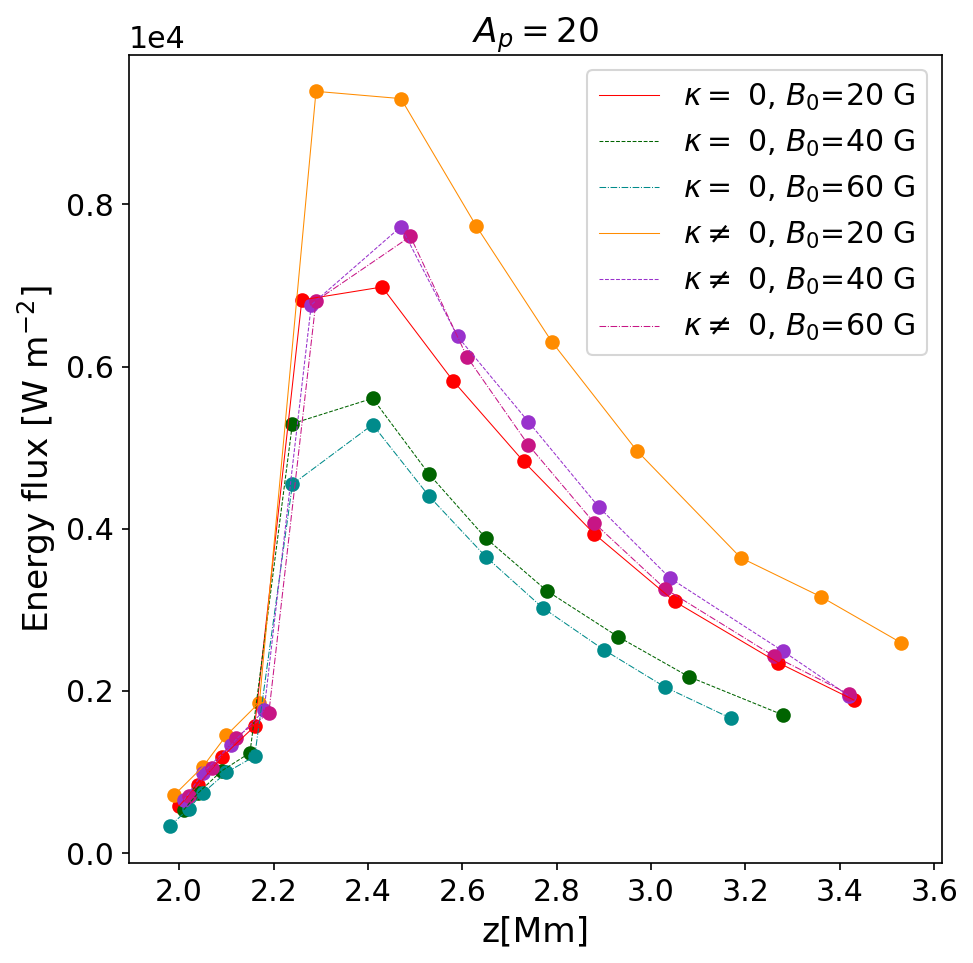}\\
\includegraphics[width = 0.32\textwidth]{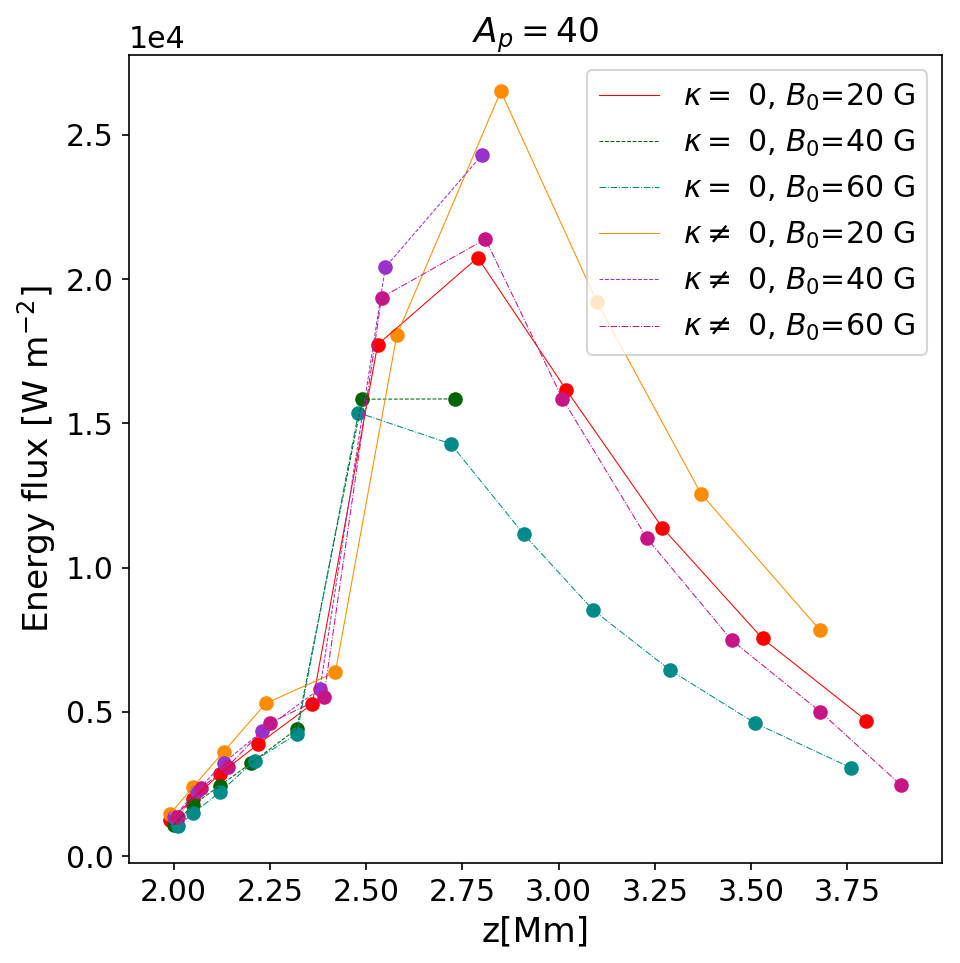}\\
\includegraphics[width = 0.32\textwidth]{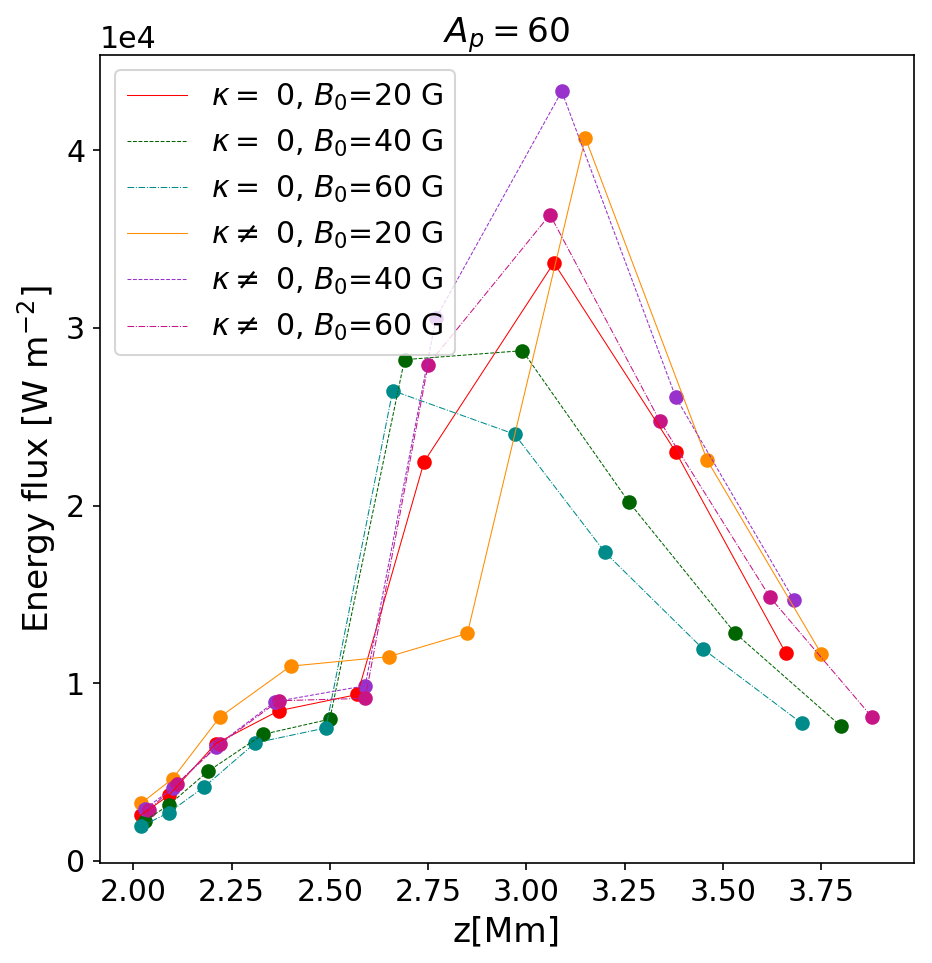} 
\caption{Energy flux in the $z$-direction carried by the jets for $A_p = $ 20, 40, 60, in the top, middle and bottom plots, respectively. \label{fig_E_flux_z} } 
\end{figure}

We compare the energy flux carried by the jet in Figure \ref{fig_E_flux_z} which displays the vertical energy flux in the vertical axis as a function of the height. The energy flux is calculated by means of the relation $\rho v_z^2 c_s$, where $c_s$ represents the sound speed \citep{2012ApJ...755...18V}. We calculate the  energy flux at the point with highest density in the jet. Each dot in the plot represents the flux energy at a different time.  Every panel corresponds to a different value of the amplitude of the gas pressure driver. In each case there is a larger energy flux in the simulations with thermal conduction denoted at the labels with $\kappa \neq 0$ in comparison to the adiabatic case  ($\kappa = 0$). In the top plot, drawn for $A_p = 20$, we found that simulations with thermal conduction the energy flux is increased from 10$\%$ to $50\%$.  The middle row is associated with $A_p = 40$. In this case the energy flux increment is up to a 50$\%$. The same trend is found in the bottom plot, which corresponds to the simulations with $A_p = 60$, the energy flux is increased up to 60$\%$. On the other hand, looking at the changes with respect to the magnetic field amplitude $B_0$, for larger $B_0$ the energy flux is smaller in the ideal regime, the energy increase was up to 35$\%$ for every value of $A_p$. Meanwhile, in the simulations with thermal conduction the increase was of $20\%$.

\begin{figure}
\centering
\includegraphics[width = 0.32\textwidth]{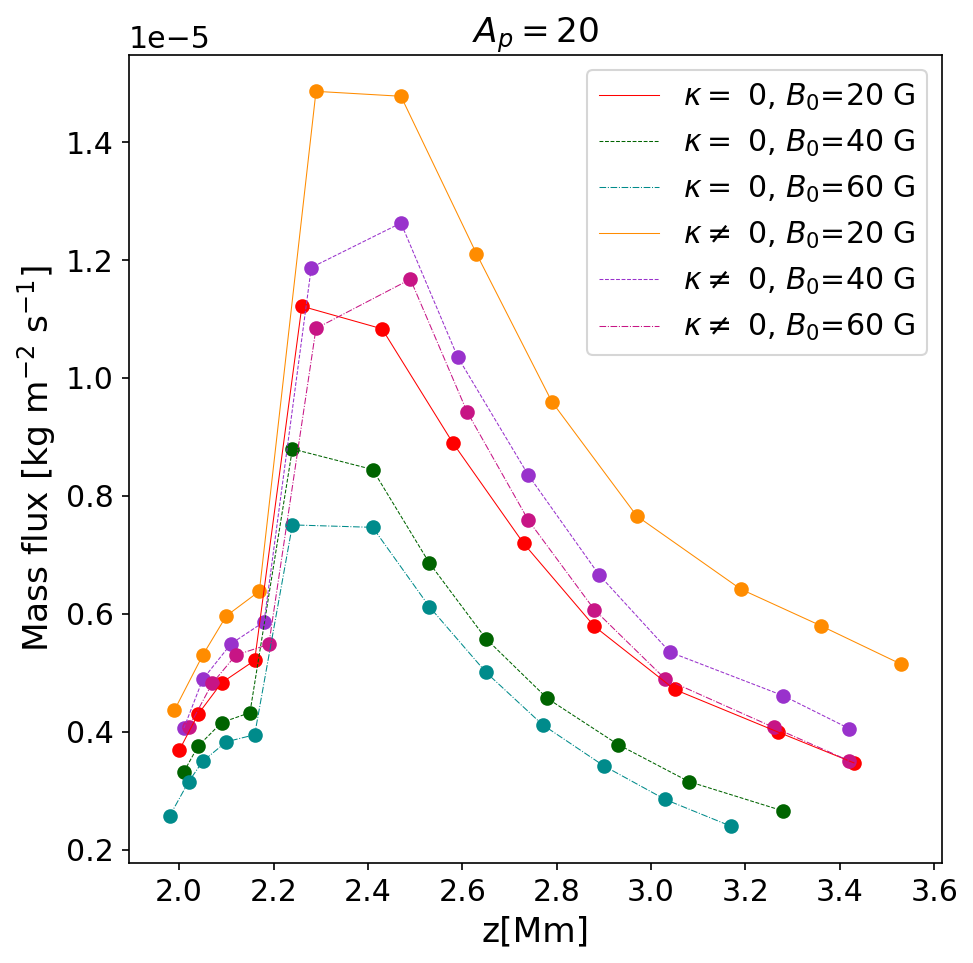}\\
\includegraphics[width = 0.32\textwidth]{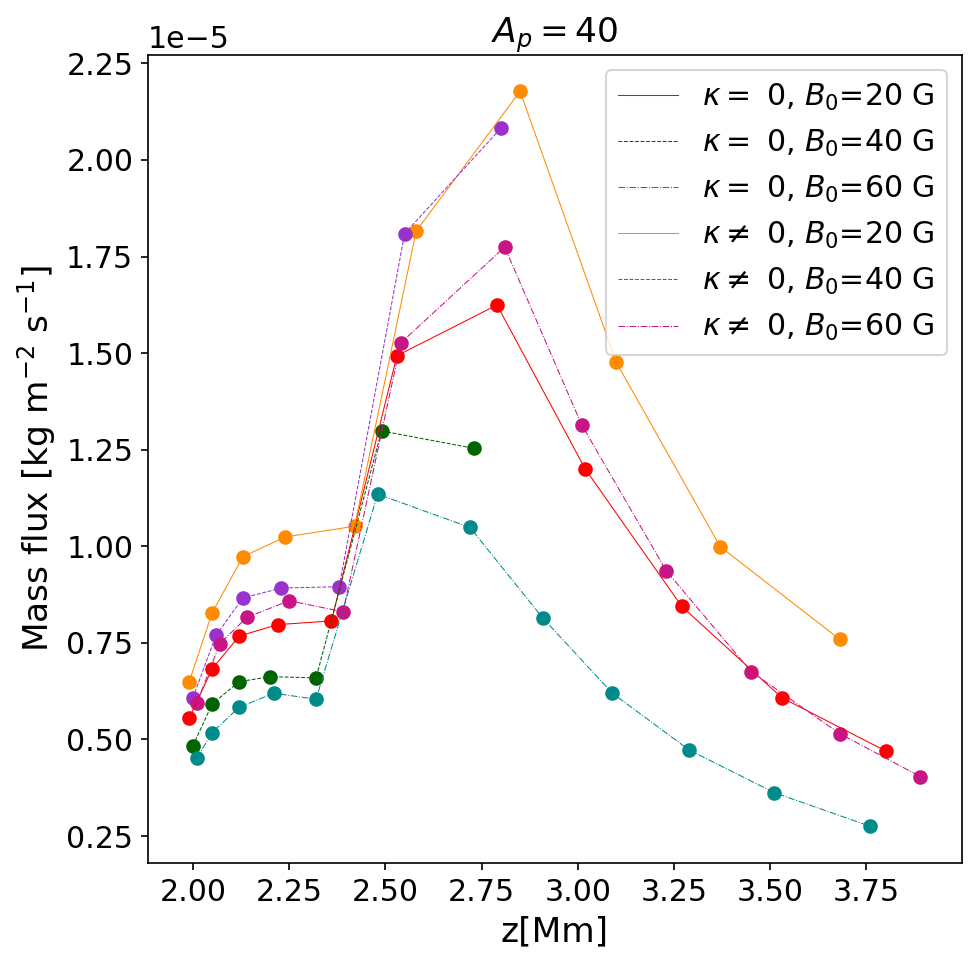}\\
\includegraphics[width = 0.32\textwidth]{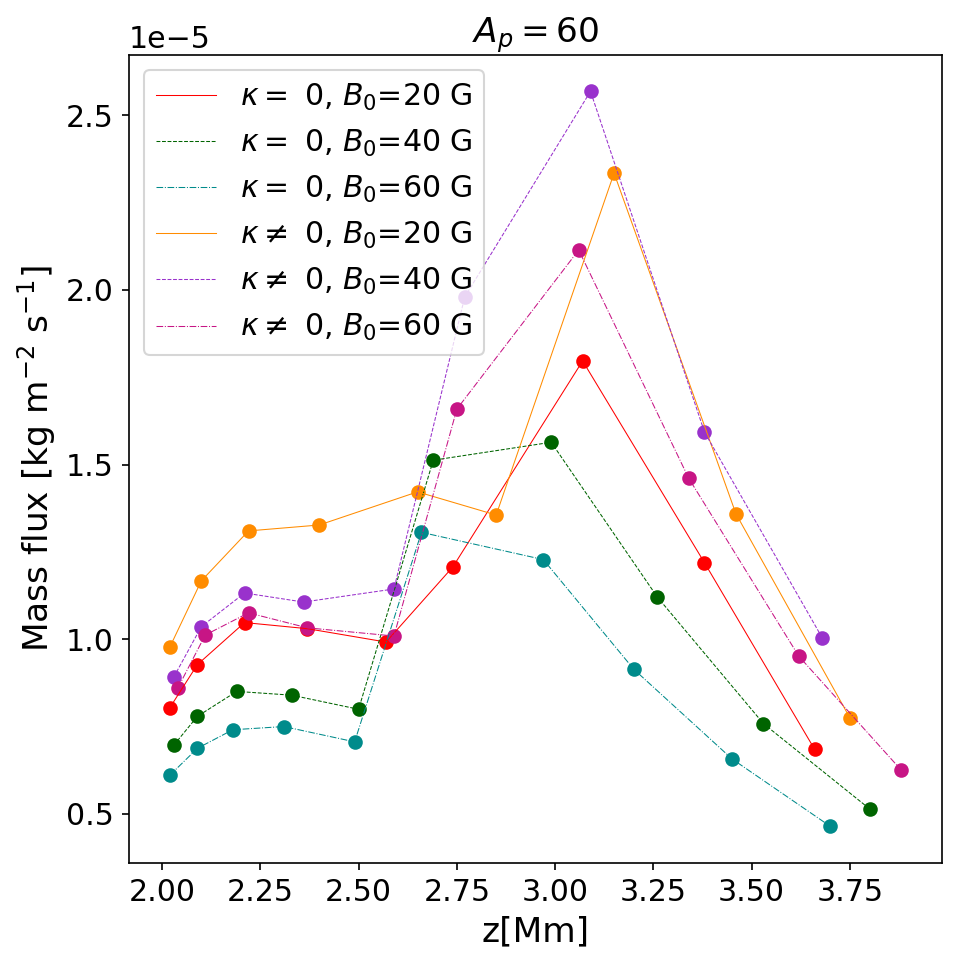}
\caption{Mass flux in the $z$-direction carried by the jet for $A_p = $ 20, 40, and 60, in the top, middle and bottom plots, respectively. \label{fig_pz} }
\end{figure}

We calculate the vertical mass flux carried by the jet and present it in Figure \ref{fig_pz} as a function of height for the different parameters in a similar fashion as in the previous plots. When the thermal conduction is included the mass flux grows from 20$\%$ to 55$\%$ for the value of the pressure amplitude $A_p = 20$. For larger amplitude drivers, mainly for $A_p = 40$ and $A_p=60$, the enhancement was from 10$\%$ to $70\%$. The mass flux with the magnetic field amplitude $B_0$ was decreased up to $35\%$ in comparison to the ideal simulations, and $30\%$ for the cases with thermal conductivity. 

\begin{figure}
\centering
\includegraphics[width = 0.33\textwidth]{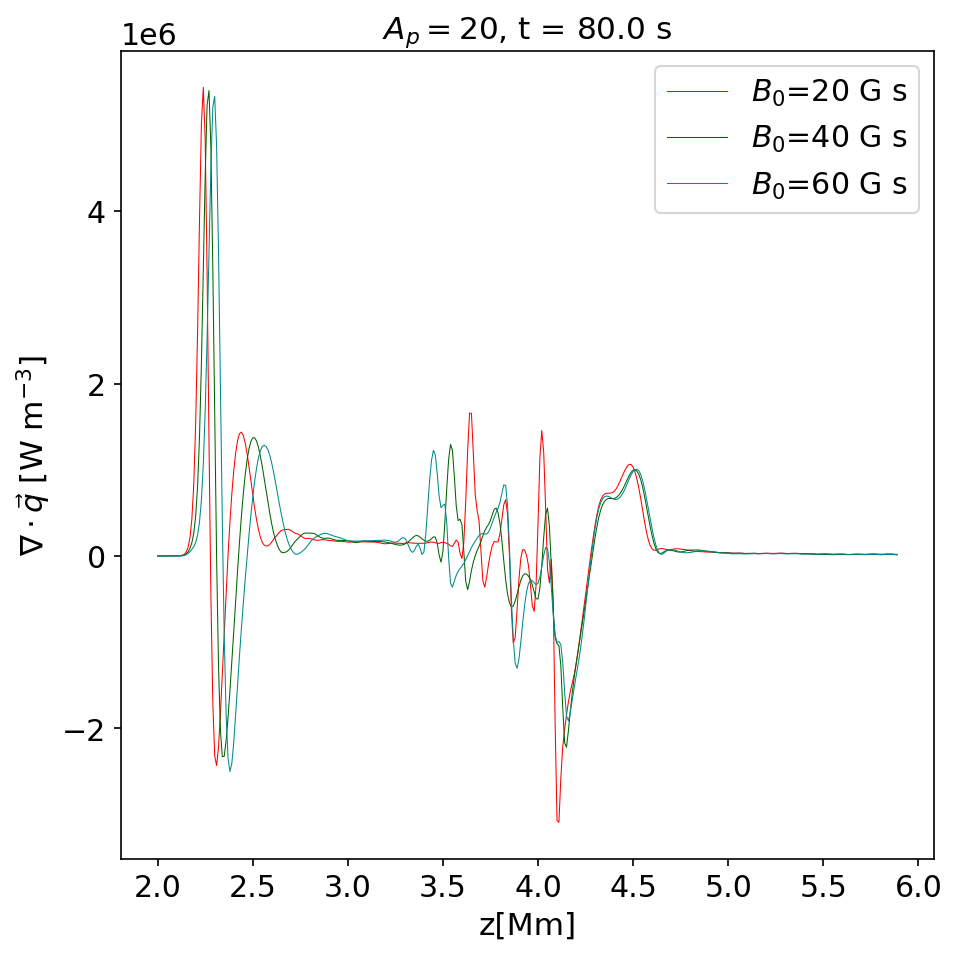}\\
\includegraphics[width = 0.33\textwidth]{heat_loss_averaged_chuncked_comparatives_time_80_Ap_20.png}\\
\includegraphics[width = 0.33\textwidth]{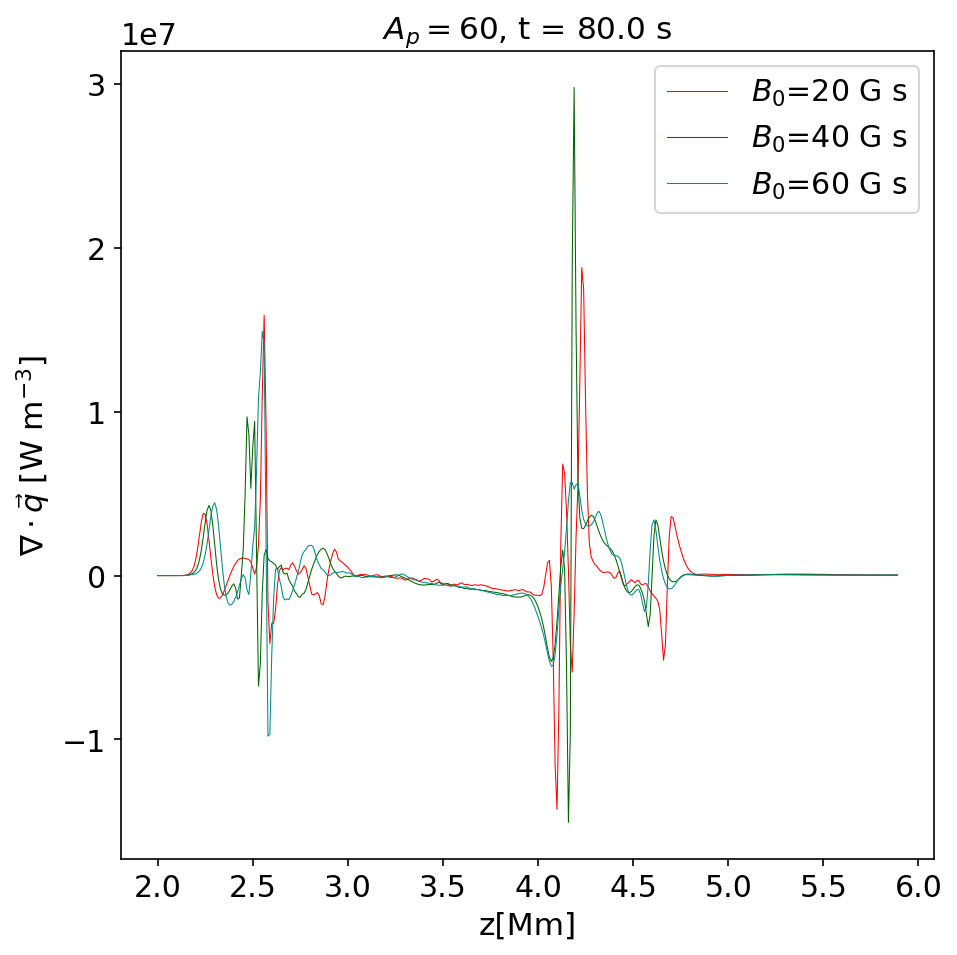}

\caption{Horizontally averaged heat loss rate at $t= 80\;$s for $A_p = 20$ (top), $A_p = 40$ (middle) and $A_p = 60$ (bottom). \label{fig_heat_loss_prom} }
\end{figure}

The thermal heating loss rate that is averaged horizontally at each height is presented in Figure \ref{fig_heat_loss_prom} at $t=80\;$s. Each panel has a different amplitude of the gas pressure driver; $A_p = 20$ at the top, the heating losses attain large values at heights $z = 2.24\;$Mm and at $z = 4\;$Mm, reaching values of $2\times 10^{6}\;$W m$^{-3}$, very similar for each value of the amplitude of magnetic field $B_0$. The middle panel corresponds to $A_p = 40$. In this case the heating loss is large at three different heights, $z = 2.5\;$Mm, $z = 3.2\;$Mm and $z = 4.5\;$Mm and attains values of the order of $1\times10^{7}\;$W m$^{-3}$. The panel at the bottom corresponds to $A_p = 60$. In this case, the order of magnitude of the heating loss is quite similar, but the peaks are shifted to $z=2.5\;$Mm and to $z=4.2\;$Mm.  

\begin{figure}
\centering
\includegraphics[width = 0.4\textwidth]{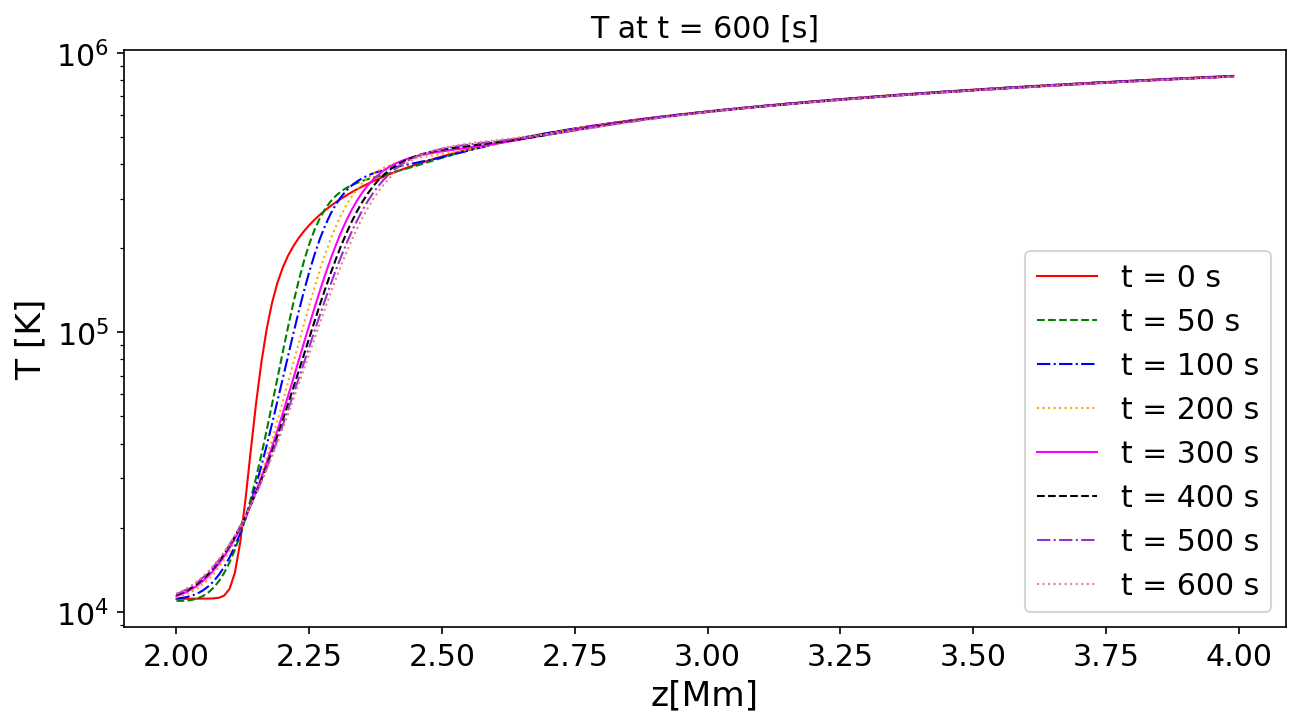}
\caption{Temperature evolution as a function of height for $t= 0, 50, 100, 200, 300, 400, 500, 600\;$s \label{fig_3} \label{fig_T_evolution} }
\end{figure}

In order to determine whether the effects observed in our simulation are created by the pressure perturbation, we simulated the evolution of the system without the pressure driver, while keeping the effects of the thermal conduction. In Figure~\ref{fig_T_evolution}, we present the evolution of the temperature as a function of height for several times between $t=0\;$s and $t=600\;$s, viz.\ at $0,50,100,200,300,400,500$, and $600\;$s. We remark that the timescales of the changes of the temperature are much larger than the timescales of our simulation including the pressure driver. Therefore, we can infer that the effects on the energy flux and mass flux mentioned in our paper are a result of the tadpole jet, since it gets around $t= 50\;$s to reach the corona. 

%%%%%%%%%%%%%%%%%%%%%%%%%%%%%%%%%%%%%%%%%%%%%%%%%%%%%%%%%%%%%%%%%%%%%%%
\section{Discussion and conclusions}
\label{sec_conclusions}
%%%%%%%%%%%%%%%%%%%%%%%%%%%%%%%%%%%%%%%%%%%%%%%%%%%%%%%%%%%%%%%%%%%%%%%

We modelled the generation of a jet at the coronal base through the implementation of a Gaussian pressure driver. Parametric studies were carried out by varying the magnetic field strength and the amplitude of the pressure driver. Another set of simulations were performed including the thermal conduction in order to determine its effects on the shape and energy transported by it. We found that the thermal conduction modifies considerably the morphology of the jet, its shape get collimated and penetrates more the solar corona. The energy and mass fluxes are increased 
when the thermal conduction is switched on. This effect may be caused by a redistribution of the internal energy triggered by the thermal conduction. For larger amplitudes of the gas pressure driver, $A_p$, these values increase, as expected since the perturbation is stronger. 
Additionally, when comparing simulations with different values of magnetic field strength (given by $B_0$) we found that for every value of $B_0$, the energy and mass fluxes increase when the thermal conduction is considered.  

%%%%%%%%%%%%%%%%%%%%%%%%%%%%%%%%%%%%%%%%%%%%%%%%%%%%%%%%%%%%%%%%%%%%%%%
\section*{Acknowledgements}
%%%%%%%%%%%%%%%%%%%%%%%%%%%%%%%%%%%%%%%%%%%%%%%%%%%%%%%%%%%%%%%%%%%%%%%
A. N. thanks COLCIENCIAS, Colombia, under the program ``Becas Doctorados Nacionales 647''. F.D.L-C was supported in part by VIE-UIS, under Grant No. 2493 and by COLCIENCIAS, Colombia, under Grant No. 8863. K.M.'s work was done within the project from the Polish Science Center (NCN) Grant No. 2017/25/B/ST9/00506.

\section*{Data Availability}
No new data were generated or analysed in support of this research.

%\bibliographystyle{mnras}
%\bibliography{bibliography}
%\bibliography{bibliography}
\end{document}